\documentclass[
    5p, 
]{elsarticle}
\biboptions{sort&compress}

\usepackage{amssymb}
\usepackage{amsmath}
\usepackage{hyperref}

\usepackage{xcolor}

\journal{Extreme Mechanics Letters}

\begin{document}

\begin{frontmatter}

\title{The Collective Snapping of a Pair of Bumping Buckled Beams}
\author[lion,amolf]{Lennard J. Kwakernaak\corref{cor}}
\ead{kwakernaak@physics.leidenuniv.nl}

\cortext[cor]{Corresponding author}

\author[bu]{Arman Guerra}

\author[bu]{Douglas P. Holmes}

\author[lion,amolf]{Martin van Hecke}

\affiliation[lion]{
    organization={Huygens-Kamerlingh Onnes Laboratory, Universiteit Leiden},
    addressline={PO Box 9504},
    city={Leiden},
    postcode={2300 RA},
    country={the Netherlands}
}
\affiliation[amolf]{
    organization={AMOLF},
    addressline={Science Park 104},
    city={Amsterdam},
    postcode={1098 XG},
    country={the Netherlands}
}

\affiliation[bu]{
    organization={Mechanical Engineering, Boston University},
    city={Boston},
    postcode={02215},
    state={MA},
    country={United States}
}


\begin{abstract}
    When a pair of parallel buckling beams of unequal width make lateral contact under increasing compression, eventually either the thin or the thick beam will snap, leading to collective motion of the beam pair. Using experiments and FEM simulations, we find that the distance $D$ between the beams selects which beam snaps first, and that the critical distance $D^*$ scales linear with the combined width of the two beams.
    To understand this behavior, we show that the  collective motion of the beams is governed by a pitchfork bifurcation that occurs at strains just below snapping.
    Specifically, we use a model of two coupled Bellini trusses to find a closed form expression for the location of this pitchfork bifurcation that captures the linear scaling of $D^*$ with beam width.
    Our work uncovers a novel elastic instability that combines buckling, snapping and contact nonlinearities. This instability underlies the packing of parallel confined beams, and can be leveraged in advanced metamaterials.
\end{abstract}

\end{frontmatter}

\section{Introduction}
Elastic instabilities govern many of the exotic properties of
mechanical metamaterials
\cite{rafsanjani2015,bertoldi2017,lubbers2017,yang2019,yang2016}.
Typically, these metamaterials consist of slender elements that go through collective buckling or snapping instabilities, causing the material to switch between two states \cite{yang2016}. However, more advanced functionalities require a sequence of reconfigurations of the material, controlled by carefully designed instabilities and nonlinearities \cite{coulais2018,kwakernaak2023,ding2022,yang2016,khajehtourian2020}.
The development of such materials thus requires an investigation into the complex instabilities mediated by interactions between multi-stable elements.

While constrained elastica have been thourougly studied, comparatively less is known
for systems of compressible beams in contact. First,
constrained elastica have proven to be a rich platform of multi-stability with strong interactions between elements. Both elastica in a potential field \cite{holmes2000}, and elastica in contact with walls \cite{holmes1999,chen2014,katz2017,manning2005} have been known to display multiple branches of stable solutions. Moreover, in systems with two elastica, the constraint between elements mediated by mutual contacts can be used as a source of interaction  \cite{chen2020,chen2022}. Second, for compressible beams, additional complications arise as
such beams buckle at finite strains \cite{bazant2010}. In addition, for thick beams, the buckling transition changes from supercritical to subcritical
\cite{magnusson2001,lubbers2017}.

We recently introduced a beam counter metamaterial which evolves sequentially, and for which contacts between compressible beams of various widths are crucial
\cite{kwakernaak2023}.
Because contacts in such systems are highly nonlinear, their analysis is divided into subcases based on the quantity and types of contact between elements. As
the number of elements in contact at any time remains small, such an
approach allows for the analysis of larger systems of many elements.

Here, we investigate the symmetry-breaking
of two unlike beams that buckle, make contact, and eventually snap.
Crucially, we consider two beams with different thicknesses leading to an asymmetry in the system; the beams buckle at different strains, and have different rigidities.
As the beam pairs are compressed, they traverse a sequence of reconfigurations.
After buckling, the beams come into contact and interact through a reciprocal constraint. The resulting system is initially stable, but at some critical compression loses stability, causing one of the beams to snap through.
Depending on whether the distance  $D$ between the beams is smaller or larger than a characteristic distance $D^*$, either beam can be selected to snap.
To study the emergence of this characteristic distance, we perform both experiments and numerical simulations for a range of beam thicknesses and distances.
Moreover, we derive an analytical framework that yields a closed-form solution for the scaling of $D^*$ that occurs in the experimental and numerical results. Our work
captures the behavior of a pair of bumping buckled beams, and can be extended
to a wide variety of scenarios where two unlike bistable elements are strongly coupled.

\begin{figure*}[!ht]
    \centering
    \includegraphics{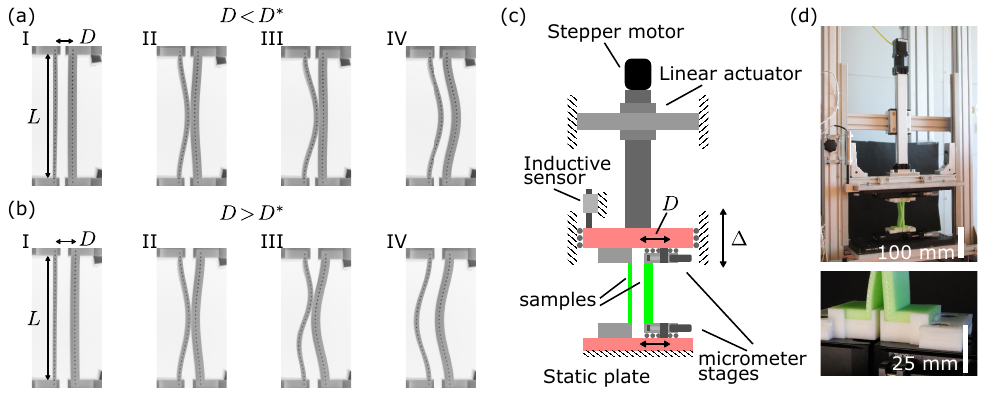}
    \caption{
        Phenomenology of two competing buckled beams.
        (a, b) Snapshots of beams separated by a distance of $D<D^*$ (a) and $D>D^*$ (b) as the compressive strain $\varepsilon$ is increased:
        (I) beams at zero strain with $D$ and $L$ indicated,
        (II) initial contact at $\varepsilon_c$,
        (III) beam configuration just before the beams lose contact at $\varepsilon^*$,
        (IV) beam configuration just after the beams have lost contact through the snapping of the thick (a) or thin (b) beam.
        (c) Diagram of the setup.
        (d) Setup used to perform experiments.
        Top: Overview of the setup.
        Bottom: Zoom in showing the beam fixture attached to the linear stage.
    }\label{fig:1}
\end{figure*}

\section{Phenomenology}

We start by discussing the qualitative nature of the evolution of
two buckling beams that come in contact under increased compression (Fig.~\ref{fig:1}a,b).
The beams have rectangular cross sections and equal lengths $L$. We non-dimensionalize all other dimensions by dividing them by $L$.
The beams are compressed by a distance $\Delta$, leading to a strain $\varepsilon=\Delta/L$.
Their out-of-plane non-dimensional thicknesses $w$ are assumed to be large and equal, so that the buckling strains are governed by the in-plane dimensionless thicknesses $t$ and $T$, where $t<T$; for definiteness, we assume that the thin beam is to the left of the thick beam (Fig.~\ref{fig:1}c).

In Fig.~\ref{fig:1}a,b we show the beam's evolution under quasistatic increase of the strain $\varepsilon$.
The thin beam buckles at $\varepsilon_t$ after which the thicker beam buckles at $\varepsilon_T$.
We assume that the beams buckle towards each other (Fig.~\ref{fig:1}aII,bII).
The distance between the centrelines of the beams, $D$, plays a crucial role, and we assume that $D$ is small enough so that the two beams eventually get into contact
at some strain $\varepsilon_c>\varepsilon_t$ --- for now we will assume that $\varepsilon_c>\varepsilon_T$ also.
When the strain is increased further, the contact forces between the beams increase, possibly leading to complex higher order mode. This configuration becomes unstable for a critical strain $\varepsilon_c$. Two distinct scenario's are then observed: either the thick beam snaps to the right (Fig.~\ref{fig:1}aIV) or the thin beam snaps to the left (Fig.~\ref{fig:1}bIV). As we will show below, the distance $D$ selects which of these two scenario's occurs, and there is a critical distance $D^*$ that separates these --- for $D<D^*$, the thick beam snaps, whereas for $D>D^*$, the thin beam snaps.
Hence, post-snapping there are two distinct states where both beams
are buckled, either to the right (Fig.~\ref{fig:1}bIV) or to the left (Fig.~\ref{fig:1}cIV).

\begin{figure*}
    \includegraphics{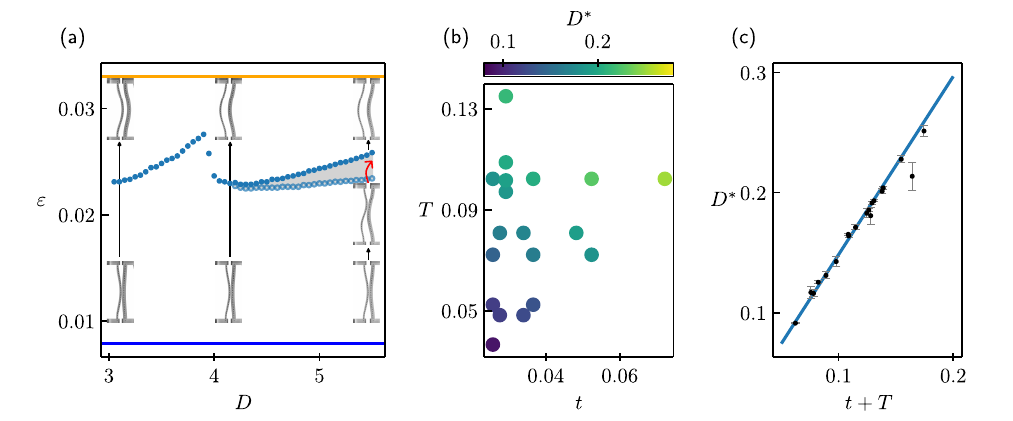}
    \caption{
        Experimental results.
        (a) Snapping strains of a two-beam pair with $t=0.026\pm0.0006$ and $T=0.072\pm0.0006$ (filled circles). Shown are results for multiple experiments where $\varepsilon$ is increased at fixed $D$.
        Insets show pictures of different stable configurations of the system. Note that at large $D$ there is an intermediary range where the thin beam becomes asymmetric before snapping.  The region where the beams are stable in the asymmetric mode is indicated in gray.
        The horizontal lines correspond to $\varepsilon =6.73t^2$ and $\varepsilon =6.73T^2$ which are the thresholds for asymmetric snapping of the thin (blue) and thick (orange) beam respectively\cite{pandey2014}.
        (b) Scatter plot of $D^*$ for $19$ combinations of $t$ and $T$.
        (c) $D^*$ plotted as a function of $t+T$ shows a simple proportional relation with a slope $\lambda^{exp}=1.484\pm 0.006$.
    }\label{fig:2}
\end{figure*}

We note that in this example,
the thick beam snapping for $D<D^*$ remains top-down symmetric (Fig.\ref{fig:1}a), while the thin beam snapping for $D>D^*$ develops an asymmetric shape (Fig.~\ref{fig:1}b).
This is consistent with the condition for the development of asymmetric beam shapes for transversely loaded buckled beams,
which according to Payndey et al. should occur at $\varepsilon=6.73t^2$ and $\varepsilon=6.73T^2$ for the thin and thick beam respectively \cite{pandey2014}.
Hence, symmetric and asymmetric snapping is determined by comparing the snapping strain of the beams, $\varepsilon_s$, with these conditions (Fig.~\ref{fig:2}).
Consistent with this, here we typically observe symmetric snapping when $D<D^*$ and asymmetric snapping when $D>D^*$, although deviations of this can occur for $T\approx t$. We note that the beam shape does not influence the left or right snapping of the beams, i.e., the value of $D^*$.

Intuitively, the emergence of the two distinct scenario's can
be understood by considering the lateral stiffnesses of the two beams as $\varepsilon$ increases.
{We define the lateral stiffness as the resistance of a beam
to a vanishingly small point load applied at the middle of a beam
perpendicular to the axis of compression. This lateral stiffness
varies non-monotonically as the beam buckles: First the stiffness decreases down to zero at the buckling point, after which it increases again in the buckled configuration.}
By taking a small enough $D$, $\varepsilon_c$ approaches $\varepsilon_T$, so that upon contact the thick beam is barely buckled and its lateral stiffness is near zero, whereas the thin beam is deeper in the post-buckling regime and significantly stiffer.
Upon further compression, the thin beam induces a snapping of the thick beam.
For even smaller $D$, $\varepsilon_c$ becomes smaller than $\varepsilon_T$. Then, as the thick beam is not yet buckled when the beams make contact, the left-right symmetry of the thicker beam is broken, determining its buckling direction rightward. 
In contrast, for large enough $D$, when the beams come into contact when both beams are significantly curved, the thicknesses of the beams dominate their lateral stiffness, and the thick beam induces snapping of the thin beam.
While intuitive, this picture does not produce a quantitative insight into what controls $D^*$, which is the focus of the remainder of this paper.

\subsection{Experimental observations}

To systematically explore the evolution of two post-buckled beams in contact, we designed and built a custom compression device which is stiff in all rotational and shear directions and ensures high parallelity between top and bottom plates
(Fig.~\ref{fig:1}c,d).
The compressive strain $\varepsilon$ is applied through a linear stage, controlled by a stepper motor and monitored with an inductive probe, yielding repeatable positioning with an accuracy of $0.05\;\text{mm}$ under typical loads. The distance $D$ between adjacent beams is controlled by four Thorlabs XRN25 manual micrometer stages housing the fixtures which hold the beams in place with an accuracy of $0.01\; \text{mm}$.
We track the deformation of the beams indicated by white protrusions on the front of the beams with a grayscale CMOS camera at a resolution of $3088\text{x}2064$, reaching a pixel density at the objective plane higher than $10\; \text{pixels/mm}$.

We studied the evolution of pairs of beams of length $L=79.8 \;\text{mm} \pm 0.05 \;\text{mm}$ and various thicknesses $t$ and $T$.
The samples studied are made out of VPS (Zhermack Elite Double 32, Young's modulus $E\approx 1\;\text{MPa}$, poisson ratio $\nu \approx 0.5$)
using molds made with FDM 3d printing on commercial UltiMaker S3 and S5 printers. After curing, the samples were allowed to rest for at least one week, well past the setting time of $22\;\text{min}$, to allow the material properties to settle \cite{florijn2016,chen2021}, after which the samples are demolded. Following this, the dimensions of the final samples were measured using an Instron universal measurement device equipped with $10\;\text{N}$ load cell and a touch probe to measure the thickness of the relatively soft beams at various locations. The standard deviation in $T$ along the surface of the samples is $0.1 \;\text{mm}$.
Experiments were only conducted with beams from the same batch of rubber.

To measure $D^*$, we performed multiple measurements for each beam pair at various $D$. At the start of each measurement, each beam is manually manipulated such that its buckled state is towards the adjacent beam. We then slowly increase $\varepsilon$ at a rate of $4.2 \times 10^{-4}\; \text{s}^{-1}$ until the beams snap.
For a typical beam pair with $t=0.026\pm0.0006$ and $T=0.072\pm0.0006$ (the same pair as in Fig.~\ref{fig:1}a), as we incrementally increase $D$ between measurements, we observe that $\varepsilon_s$ varies smoothly up until $D\approx0.153$, as can be seen in (Fig.~\ref{fig:2}a).
Here $\varepsilon_s$ sharply decreases as the system both transitions from displaying the below-$D^*$ to above-$D^*$ phenomenology, as well as shifting from a symmetric snap-through mode to an asymmetric snap-through mode.
We note that the transition between left and right snapping, and the transition between symmetric and asymmetric beam shapes, are independent.
The transition from symmetric to asymmetric beam shapes is determined solely by
the values of  $\varepsilon/t^2$ and $\varepsilon/T^2$   --- for the example here, $6.73 t^2 < \varepsilon_s < 6.73 T^2 $ \cite{pandey2014}, so that the thick beam remains symmetric while the thin beam takes on an asymmetric shape (Fig.~\ref{fig:2}a).
Finally, we observe that as $D$ is increased above $D^*$, a small strain range opens up where the asymmetric beam shape is stable, before snapping at a larger strain (Fig.~\ref{fig:2}a).

Monotonously increasing $D$ such as in Fig.~\ref{fig:2}~a, unintentionally trains the samples, such that the apparent value for $D^*$ differs for increasing and decreasing sweeps of $D$.
To minimize this hysteresis and accurately measure $D^*$, we performed iterative measurements with a specific protocol that reduces the number of subsequent measurements above and below $D^*$.
We chose initial large steps of $L\cdot \left(D_{i+1}-D_{i} \right) = 1\; \text{mm}$ to find bounds on $D^*$, and then refined the bounds with decreasing stepsizes: $0.5\; \text{mm}$, $0.25 \;\text{mm}$, $0.1\;\text{mm}$ and finally $0.05\;\text{mm}$.
We then repeated every measurement set with exchanged left and right beams to correct for small asymmetries in the setup. We finally estimate $D^*$ and calculate an error through the average and RMS of the four measured bounds.

Our experiments yield $D^*$ for nineteen pairs of beams (Fig.~\ref{fig:2}b). We note that $D^*$ grows with both $t$ and $T$, and surprisingly, the data for $D^*$ can be collapsed on a single axis by plotting it as a linear function: $D^* = \lambda^{exp} (t+T)$ (Fig.~\ref{fig:2}c), with a least squares fit slope of $\lambda^{exp} = 1.484\pm 0.006$. We note that this data collapse does not significantly improve by adding an empirical fit parameter $\ell$, i.e. plotting $D^*$ as a function of $t+\ell T$.
We discuss the validity and underlying physics that leads to this collapse in section~\ref{sec:modeling}.

\subsection{Finite Element Simulations}

\begin{figure}
    \includegraphics{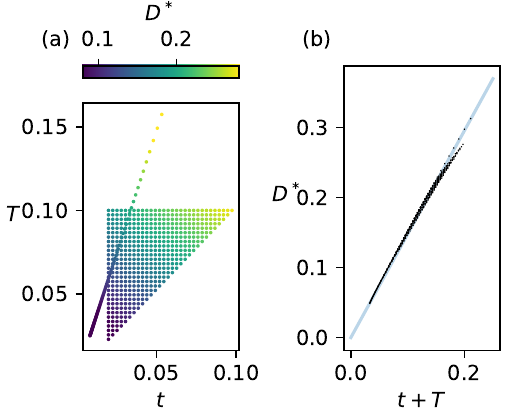}
    \caption{
        Finite element data for $D^*$ as a function of $t$ and $T$. (a) Raw data, showing the range of parameters of our two sets of
        simulations. (b) The data for $D^*$ collapses as a function of $t+T$. The line corresponds to a linear fit of the data for $0 < t+T < \frac14$, where $\lambda^{fem}=1.478\pm0.002$. Note that the density of points is not uniform along the $t+T$ axis.
    }\label{fig:abaqus}
\end{figure}

To eliminate the role of plasticity and to test the validity of our observations for a  wide variety of beam parameters, we performed FEM simulations of the co-buckling beams using ABAQUS with explicit time-stepping, CPS4 elements, Neo-Hookean material properties with a Poisson ratio of 0.49, uniform element sizes and sufficient damping to prevent oscillations.
To ensure the beams buckle towards each other, a small temporary load is applied before the beams buckle and removed before the beams make contact.
We performed two sets of simulations.
In the first set we varied both $t$ and $T$ between $0.01$ and $0.1$,
while in the second we varied the length of the beams at constant ratio $T/t \approx 2.95 $.
For every parameter $t$ and $T$, we performed multiple simulations using a bisective approach to determine $D^*$, until the error in $D^*$ was less than $10^{-4}$.
The results of these simulations are shown in (Fig.~\ref{fig:abaqus}).

Similar to our experiments, we found both symmetric and asymmetric snapping.
Consistent with our experimental observations, we find that $D^*$ is essentially proportional to $t+T$ for $t+T<0.15$.
The fit of the numerical data yielded the slope: $\lambda^{fem}=1.478\pm0.002$, which is consistent with the results of the experimental data where $\lambda^{exp}=1.484\pm 0.006$.
We conclude that the critical distance $D^*$ is linear in $t+T$.

\begin{figure*}
    \centering
    \includegraphics{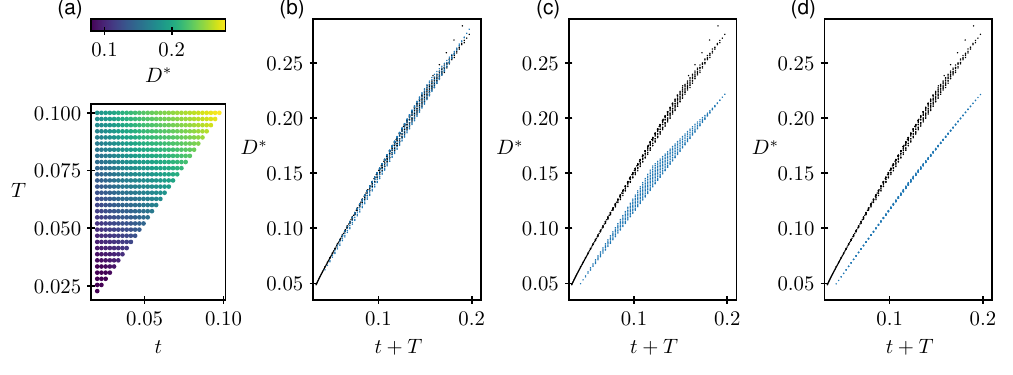}
    \caption{
        The critical distance $D^*$ obtained in the elastic bead-chain model.
    (a) Scatter plot of the calculated $D^*$ for $N = 62$. (b-d)
    $D^*$ collapses when plotted as a function of $t+T$. Data for the bead-chains and FEM simulations in blue and black respectively ((b) $N = 62$ ; (c) $N = 4$ and (d) $N = 2$.).
    }\label{fig:lammps}
\end{figure*}

\section{Simplified models and theory}
\label{sec:modeling}
The phenomenology of the joint snapping of buckled beam pairs hints at the existence of a pitchfork bifurcation that occurs when the beams are in contact, i.e., before the beams snap through.
Here we ask what the minimal ingredients are to observe such a pitchfork scenario. First, we investigate joint snapping for a slender beam model consisting of spherical beads connected with $N$ bars that are modeled as linear and torsional springs, as proposed by Guerra et. al. \cite{guerra2023}. We find that for large $N$, this simplified model captures the full phenomenology, including the existence of $D^*$ and both symmetric and asymmetric snapping.
For decreasing values of $N$, the model becomes more crude, but the existence and linear relation of $D^*$ with $t+T$ remains valid down to $N=2$. Such $N=2$ beams, which we call Bellini trusses \cite{bellini1972}, clearly cannot have asymmetric shapes, again indicating that asymmetry is not essential for the understanding of the scaling of $D^*$.
Second, inspired by these empirical observations, we study the joint buckling and snapping of pairs of Bellini trusses in section \ref{seq:Bellinitruss}. We show that their left or rightward snapping does not require the beams to lose contact, allowing us to focus on pairs of connected Bellini trusses. Finally, we show that the joint buckling and snapping is an example of a general scenario involving pairs of interacting elements that undergo pitchfork bifurcations at different values of the control parameter $\varepsilon$.
We expand the Bellini truss system to analytically solve for $D^*$ and find that it is linear in $t+T$ (in lowest order).
Together, this shows that joint snapping and the emergence of $D^*$ is a robust and universal phenomena.

\subsection{Elastic Bead-Chains in Contact}
We model the contact dynamics of post-buckled beams with a simplified model
of hard beads connected by Hookean and torsional springs.
For a large number of links $N$, this model has been shown to accurately and computationally effectively model the dynamics of collections of buckled beams in contact \cite{guerra2021,guerra2023}.
In addition, in the limit of small $N$ ($N=2$), the model converges to an initially straight Bellini truss \cite{bellini1972}.
In the beam-chain model we space our nodes equidistantly along the beam length and choose the spring constants to match the stretching and bending energy of realistic beams \cite{guerra2021}. We implement the contact dynamics between the beams with a stiff Hertzian contact model. The ends of the beams are controlled through the top and bottom particles, which control $\varepsilon$ and $D$ and which enforce the "fixed-fixed" boundary conditions of the beams (for details see SI).
We implemented the model beams using damped explicit time-stepping with the Large-scale Atomic/Molecular Massively Parallel Simulator (LAMMPS) \cite{thompson2022,guerra2023}.

We performed simulations of pairs of beams for: $N=62$, $N=4$ and $N=2$, using the same bisective protocol to determine $D^*$ (Fig.~\ref{fig:lammps}). Here $N=62$ approximates the continuous limit, while $N=2$ corresponds to the smallest possible number of segments.
For $N=62$, the characteristic linear scaling $D^*=\lambda^{md}_{N=62} (t+T)$ emerges with $\lambda^{md}_{N=62}=1.472\pm0.002$, consistent with both the experimental and finite element simulation data (Fig.~\ref{fig:lammps}b).
We note that these simulations also capture the symmetric and asymmetric beam shapes.
Strikingly, for $N=4$ and $N=2$ (where the shape is purely symmetric)
a comparable linear scaling of $D^*\propto t+T$ occurs ($\lambda^{md}_{N=4}=1.182\pm0.002$ and $\lambda^{md}_{N=2}=1.1542\pm0.0007$) (Fig.~\ref{fig:lammps}(c-d)).

\begin{figure*}
    \centering
    \includegraphics{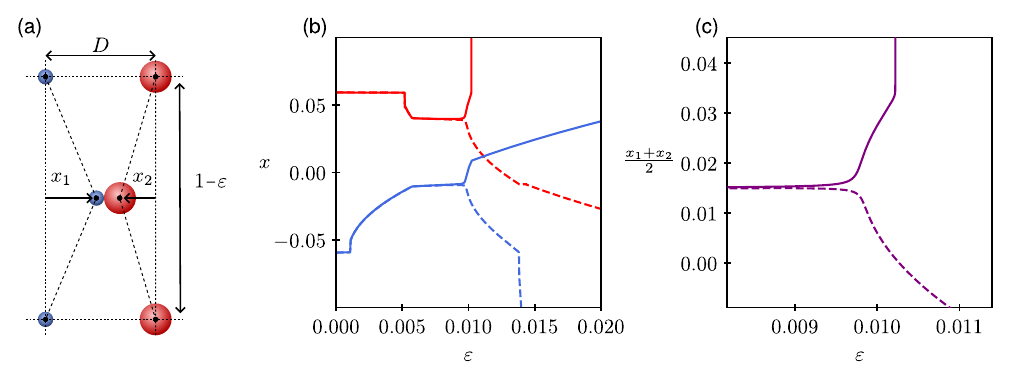}
    \caption{
        Pitchfork bifurcation for a pair of Bellini trusses.
        (a) The geometry for the $N=2$ system indicating the transverse displacements $x_1$ and $x_2$ for at a strain $\varepsilon$.
        (b) The horizontal positions $x_1,x_2$ (red and blue respectively) of the middle nodes of the two trusses for $N = 2$ as function of strain $\varepsilon$, for two  values of $D$ just below (dotted) and just above (full) $D^*$ ($t \approx 0.0031$, $T \approx 0.0071$, $|D-D^*| \approx 10^{-6}$).
        (c) The mean horizontal position $\langle x \rangle := (x_1+x_2)/2$ near the pitchfork bifurcation point of these two cases track each other closely until they branch of at $\varepsilon \approx 0.0095$.
    }\label{fig:lammps-n2}
\end{figure*}

For all three cases, we note that the beams first collectively move left or right, and then snap at a higher value of $\varepsilon$. This is illustrated in Fig.~\ref{fig:lammps-n2} for the simplified $N=2$ case, where we compare the evolution of the lateral motion of the center nodes as function of the strain for two values of $D$ just above and below $D^*$. Our data strongly suggest that the $D=D^*$ case correspond to a pitchfork bifurcation, with the evolution for $D\ne D^*$ given by unfolding of this pitchfork bifurcation. At larger strain, the discontinuous snapping transition occurs, but to determine $D^*$, it suffices to determine the location of the pitchfork bifurcation.

\subsection{Instabilities in a pair of Bellini trusses}
\label{seq:Bellinitruss}
To understand the mechanisms that govern the critical distance and its scaling with $t+T$, we analytically determine the critical values $(D^*, \varepsilon^*)$
of the pitchfork bifurcation
in the model based on a pair of initially straight Bellini trusses.
First, we connect their center nodes to model the persistent contact near the bifurcation, and separate the end nodes to capture $D$ and $\varepsilon$ (See Fig.~\ref{fig:lammps-n2}a).
Specifically, we place the end points of the thin and thick beams at $x=\alpha$ and $x=\beta$, and require that
\begin{equation}
    D = \beta - \alpha + \frac t2 + \frac T2
    \text{,}
    \label{eq:distance}
\end{equation}
where we account for the thickness of the beams.

Second, we expand the elastic energy of both Bellini trusses up to quartic order in $x$ and linear order in $\varepsilon$, and and find at leading order (see SI):
\begin{equation}
    U_t=    (\xi t^2 - \varepsilon) tx^2 + t x^4
    \text{,}
\end{equation}
where $\xi t^2$ is the buckling strain with $\xi=\frac{4B}{K}$. Hence, the buckling strain scales as the ratio of the constants $B$
and $K $
which parametrizes the compressive stiffness $Kt$ and bending stiffness $B t^3$ in the truss. Here,
$\xi$ can be considered  the inhibition to buckling due to the applied boundary conditions and degrees of freedom of the beam model (see SI).

Satisfying Eq.~\ref{eq:distance}, we obtain the total potential energy:
\begin{multline}
    U =   t\left[(\xi t^2-\varepsilon)(x-\alpha)^2 + (x-\alpha)^4\right] + \\ T \left[(\xi T^2-\varepsilon)(x-\beta)^2 + (x-\beta)^4\right]
    \text{.}
\end{multline}

We now obtain a closed form expression for $(D^*, \varepsilon^*)$ by locating the pitchfork bifurcation in this quartic energy expansion. We first, without loss of generality, choose $\alpha t + \beta T =0 \to \alpha = -\beta \frac Tt$ to eliminate the cubic terms in the expansion. Hence, $D=\beta (1+\frac T t)+\frac t 2 + \frac T 2$ and we then write the potential in the form:
\begin{equation}
    U = U_0 + a x + b x^2 + c x^4
    \text{.}
\end{equation}

The stable and unstable equilibria of the system are found at the roots of $F = \frac\partial{\partial x} U$, where:
\begin{align}
    F &= a + 2 b x + 4 c x^3 \\
    &= c\cdot( q + p x + x^3 )
    ~\text{,}
\end{align}
with:
\begin{equation}
    q = q(\beta)= \frac{2T^3\beta^3 - 2T\beta^3t^2 - T^3\beta\xi t^2 + T\beta\xi t^4}{2Tt^2+2t^3}
    \text{,}
    \label{eq:q_expr}
\end{equation}
\begin{equation}
    p = p(\varepsilon, \beta) = \frac{6T^2\beta^2 + 6T\beta^2t-T\varepsilon t+T^3\xi t-\varepsilon t^2+\xi t^4}{2Tt+2t^2}
    \text{.}
    \label{eq:p_expr}
\end{equation}

Crucially, we do not need to solve for the roots of $F$ explicitly; to find the bifurcation point, we only need to detect a change in the number of roots.
The multiplicity of the roots of $F$ can be determined from the discriminant $\Delta\{F/c\}=4 p^3 + 27 q^2$.
We note that this strategy is generally applicable for polynomials of arbitrary degree, whereas finding the solutions to such polynomials is generally not possible.
As $\varepsilon$ increases, the system changes from monostability to bistability.
For $D=D^*$, this happens through a pitchfork bifurcation at $\varepsilon=\varepsilon^*$. For $D\neq D^*$ this happens through a saddle node bifurcation.
This change of stability corresponds to $\Delta\{F/c\}$ crossing $0$,
where the pitchfork bifurcation occurs for $q=0$ and the saddle node bifurcation occurs otherwise; in the latter case, the location of the saddle node determines whether the beams move left or right.
As $p$ depends only on $\beta$ and not $\varepsilon$ (Eq.~\ref{eq:p_expr}), we can solve for $\beta^*$:
\begin{equation}
    \beta^*=t \sqrt{\frac \xi 2}
    \text{,}
    \label{eq:bstar_solution}
\end{equation}
which can be substituted into Eq.~\ref{eq:distance} to obtain $D^*$:
\begin{equation}
    D^*=(t+T)\left(\sqrt{\frac \xi 2} + \frac12\right)
    \text{.}
    \label{eq:dstarexpr}
\end{equation}
In addition, we obtain the critical strain by solving $q=0$
at $\beta=\beta^*$ and obtain
\begin{equation}
    \varepsilon^* = \xi (t+T)^2
    \text{.}
\end{equation}

We thus find that $D^*$ scales linearly with $t+T$, consistent with our experimental and numerical results. In addition, we find a testable relation between the slope $\lambda$ and the
strain at which the beams buckle, as both depend on $\xi$:
$\lambda=\frac12 + \sqrt{\frac \xi 2}$, while $\varepsilon_t=\xi t^2$.
Thus we predict that the boundary conditions of the beams influence $D^*$, e.g. pinned-pinned beams will have a smaller $D^*$ than fixed-fixed beams.
Comparing the Bellini truss model to the $N=2$ simulations with $\xi=\frac{3}{4}$ (see SI), we find a predicted $\lambda=\frac12 +\frac9{16} = 1.0625$, which is comparable to the value obtained from simulations: $\lambda^{md}_{N=2}=1.1542\pm0.0007$.

\section{Conclusion and discussion}

We studied the collective snapping of two buckled beams in contact
by means of experiment, numerics and theory.
Using experiments and FEM simulations, we found a linear relation between the critical distance and the combined width of the two beams: $D^*=\lambda(t+T)$.
We studied a simplified model consisting of $N$ compressive  rods connected by
torsional springs \cite{guerra2023}.
We find that
at large $N$, this model accurately captured the collective snapping and critical distance, while at small $N=2$, the model allows to identify
the essential mechanism that controls the eventual direction of snapping:
a pitchfork bifurcation that occurs at critical strain $\varepsilon^*$ and distance $D^*$. Furthermore, this model allows to obtain a closed form solution for $\varepsilon^*$ and $D^*$ which captures the linear relation between $D^*$ and $t+T$.

Our approach can be extended to a wide variety of scenarios where two bistable elements are strongly coupled, e.g., where the collective state can be described by a single coordinate. These include Bellini trusses that are precurved, and more generally, any buckling elements. The essential physics is that when two systems that undergo symmetric or asymmetric pitchfork bifurcations are coupled, the collective behavior is governed by a new pitchfork bifurcation.

\bibliographystyle{elsarticle-num}
\bibliography{BattlingBeams.bib}

\cleardoublepage
\appendix

\setcounter{figure}{0}

\section{Mathematical Derivation Bellini Truss}
\label{seq:QuarticExpansionBT}

\begin{figure}
\centering
    \includegraphics{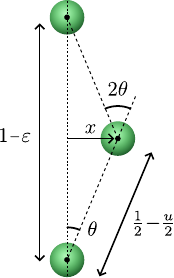}
    \caption{
        The geometry of the top-down symmetric Bellini truss.
    }\label{fig:appendix}
\end{figure}

A compressible beam can be modelled by a collection of compressive trusses serially linked by torsional springs, as shown by Guerra et.al. \cite{guerra2021}. For small angles $\theta_j$ we can write the potential of such a beam as (Fig.~\ref{fig:appendix}):
\begin{equation}
    U = \frac12 k \sum_{i}^N u_i^2 + \frac12 b \sum_{j}^{N+1} \theta_j^2
    \text{,}
\end{equation}
where $u_i$ is the compression of each spring and $\theta_j$ the change in angle from the resting configuration.
The spring constants $k$ and $b$ are chosen to match the compressive stiffness and bending stiffness of beams with a rectangular cross section, so that $k\propto t$ and $b\propto t^3$.

A Bellini truss corresponds to the
$N=2$ case.
Under imposed top-down symmetry ($u_1=u_2$ and $\theta_2=2\theta_1=2\theta_3$)
 the summation can be performed to obtain:
\begin{equation}
    U = K t u^2 + B t^3 \theta^2
    \label{eq:general_af_buckling_potential}
    \text{,}
\end{equation}
where we absorbed
the summation over the number of springs into the coefficients $K$ and $B$.

To express the potential in $x$ and $\varepsilon$, we
express $ u = 1 - \sqrt{ x^2 + \left(\frac12- \frac\varepsilon 2 \right)^2}$, $\theta = \arctan{\frac{2x}{1-\varepsilon}} $ to obtain:

\begin{equation}
    U = K t \left( 1 - \sqrt{ x^2 + \left(\frac12- \frac\varepsilon 2 \right)^2} \right)^2 + B t^3 \left( \arctan{\frac{2x}{1-\varepsilon}} \right)^2
    \text{.} \label{eq:geometric_potential}
\end{equation}

Instead of attempting to minimize the full energy Eq.~(\ref{eq:geometric_potential}), we expand it to fourth order in $x$ and first order in $\varepsilon$ around $(x,\varepsilon) = (0, 0)$, and obtain:

\begin{equation}
    U \approx (K t - \frac{32B t^3}{3}) x^4 + (4B t^3 + 8B t^3 \varepsilon - K t \varepsilon) x^2
    \text{,}
\end{equation}

As $x$, $\varepsilon$, and $t$ are all small, we discard the highest order terms $\mathcal{O}(t^3 x^4)$ and $\mathcal{O}(t \varepsilon x^2)$, and obtain the
leading order potential:
\begin{equation}
    U\approx Ktx^4 + (4Bt^2-K\varepsilon)tx^2
    \text{.}
\end{equation}
This potential transitions from a monostable to a bistable form when the $x^2$ term switches sign at $\varepsilon=\frac{4B}{K}$.
Diving by $K$ produces the rescaled potential
that makes this transition explicit:
\begin{equation}
    U = tx^4 + (\xi t^2 - \varepsilon)tx^2
    \text{,}
\end{equation}
where the transition from a monostable to bistable potential occurs  at $\varepsilon=\xi t^2$.

\end{document}